\documentclass[%
 reprint,
 amsmath,amssymb,
 aps,
floatfix,
]{revtex4-2}

\usepackage{graphicx}% Include figure files
\usepackage[T1]{fontenc}
\usepackage{comment}
\usepackage{braket}
\usepackage{float}
\usepackage{fixltx2e}
\usepackage{dcolumn}% Align table columns on decimal point
\usepackage{bm}% bold math
\begin{document}

\preprint{APS/123-QED}

\title{Channel induced dynamics of quantum information\\ in mixed state free QFTs }% Force line breaks with \\

\author{Michal Baczyk}
% \email{michal.baczyk123@gmail.com}

\affiliation{%
Faculty of Physics, University of Warsaw, ulica Pasteura 5, 02-093 Warsaw, Poland 
}%

%\affiliation{
%\affiliation{%


%\date{}% It is always \today, today,

\begin{abstract}
\noindent We propose a new framework for Quantum Field Theory (QFT) studies that allows us to represent field excitations as quantum channels. We demonstrate the inner workings of the proposed scheme for three universal states: a regularized vacuum state, a thermal state of a one-dimensional QFT system, and the lattice-regulated Thermofield Double State of two identical free QFTs. We investigate the actions of unitary and non-unitary Bosonic Gaussian channels (including Petz Recovery maps). To evaluate and quantify the character of the channel static action and channel-induced dynamics, we calculate quantum entropies and fidelities.
\end{abstract}

\maketitle

%\tableofcontents

\section{\label{sec:Introduction}Introduction}

Gaussian states, although elements of infinite-dimensional Hilbert spaces, can be expressed in terms of finite-dimensional mathematical objects: a covariance matrix and a vector of first moments that together contain all the information about the system \cite{Adesso_2014,RevModPhys.84.621}. Such a description of complex systems in terms of numerable quantities allows us to study universal quantum information features of states that possess direct relevance for holography. This valuable connection between Gaussian Quantum Information (QI) and holography was already explored in \cite{Chapman_2019}, in which the authors simulated the growth of entanglement entropy and complexity for the Thermofield Double State (TFD).

To analyze the dynamics in Quantum Field Theories (QFTs), as well as in holography, it is
necessary to consider excitations. These have been
investigated in the context of quantum quenches in Conformal Field Theories (CFTs) \cite{Calabrese:2005in, Calabrese:2016xau} and their holographic duals \cite{Maldacena_2003,Nozaki:2013wia, holography8,holography9}. From the QI perspective, we can treat excitations as quantum channels. Accordingly, the main objective of this work is to take such an operational approach and examine situations where quantum channels are applied to regularized free QFTs.

We implement our idea for three specific cases: a vacuum state, a thermal state of QFT, and a TFD state of two identical copies of QFT. We regularize these states on a lattice and cast them into the framework of Bosonic Gaussian QI \footnote{We further distinguish between Bosonic and Fermionic Gaussian Quantum Information. In this paper, we solely concentrate on the Bosonic Gaussian Quantum Information Theory.}. We can then apply various Bosonic Gaussian channels to systems under consideration. We put special emphasis on Petz Recovery maps \cite{Petz1,PETZ_2003,PETZ2}, which play a significant role in the entanglement wedge reconstruction and in resolutions of the black hole information paradox \cite{Cotler:2017erl,penington2020entanglement, Chen:2021lnq}. To determine the effect of quantum channels, we calculate von Neumann entropy, R\'enyi entropy, and fidelity. We observe how they change after the application of a channel in both time-independent and time-dependent settings. By interpreting these results, we hope to understand excitations in QFTs from the QI operational perspective.

The outline of the work is presented below.

The Methods and Methodology Section \ref{methods} introduces all the necessary notions, concepts, and tools that brought together constitute a novel framework established in this work. The introduction to Bosonic Gaussian channels (Subsection \ref{sec:channels}), description of used states and their evolution (Subsection \ref{sec:investigatedstates}), channel application for field theories in the continuum limit (Subsection \ref{sec:standardization}), as well as simulations technical summary (Subsection \ref{sec:technicalsummary}) are all included here.

The Results Section \ref{results} provides details of conducted numerical experiments that validate the presented framework and explain the features of it both in the time independent (Subsection \ref{sec:static}) as well as in time-dependent settings (Subsection \ref{sec:QIdynamics}).

The Discussion Section \ref{discussion} puts our work in the context of previous findings and highlights new insights (Subsection \ref{relatedworkandresults}). It offers further context about particular selections of boundary conditions (See Subsection \ref{section:boundary}) and the standardization process (Refer to Subsection \ref{sec:standarizationdiscussion}). Moreover, it also establishes the first step towards an extension of the existing framework to an even more generalized approach (Subsection \ref{appendix:gaussian}).

Finally, we conclude all the developments and findings in Section \ref{conclusions}.

\section{Methods and methodology} \label{methods}

To start, we explain the fundamentals of a Bosonic Gaussian QI framework and demonstrate what types of channels can be applied and how they can be applied (see Section \ref{sec:channels}). As a next step, we introduce the three states: a vacuum state and a thermal of QFT and a TFD state of two identical copies of QFT that we study throughout the paper -- we start from the field theory level and go all the way to the discretization of the states and describing them in the Bosonic Gaussian QI language (see Subsection \ref{sec:investigatedstates}). Afterwards, we introduce a necessary standardization procedure that allows us to investigate channel actions in the continuum limit (see Subsection \ref{sec:standardization}). For completeness, we provide a technical summary (see Subsection \ref{sec:technicalsummary}) of all the steps needed to achieve the results to follow in Section \ref{results}.

\subsection{Channels in Bosonic Gaussian Quantum Information framework} \label{sec:channels}

To begin with, we explain fundamentals of a Bosonic Quassian QI framework with special emphasis on channels formalism. We introduce the notation and notions of a covariance matrix and a vector of first moments, then we provide details how channel operation alters the state of a quantum system represented in the studied framework outlining a few examples of unitary and non-unitary quantum channels. Last but not least, we explain how to describe subsystems and operations acting locally.\\

Consider a $n$ bosons quantum system, for which we define a \textit{canonical operators vector}:

\begin{equation}
r = (\hat{x}_1, \hat{p}_1, \hat{x}_2, \hat{p}_2, .... , \hat{x}_n, \hat{p}_n),
\end{equation}

\noindent where $\hat{x}_i$ and $\hat{p}_i$ are canonical operators associated with boson $i$.

Then, in the unit system we employ ($ \hbar= c = k_B = 1$), the commutation relations can be expressed as \newline $\left[ r, r^{T}\right]=i J$,  where:

\begin{equation}
J=\bigoplus_{j=1}^{n} J_{1}, \quad \text { with } \quad J_{1}=\left(\begin{array}{cc}
0 & 1 \\
-1 & 0
\end{array}\right).
\end{equation}

As a consequence of Wick's theorem \cite{hackl2021bosonic}, if a quantum state with a density matrix $\rho$ is Gaussian, then it is fully characterized by its \textit{vector of first moments} $s_{\rho}$ and its \textit{covariance matrix} $V_{\rho}$ defined as follows:

\begin{equation} \label{eq_gauss1}
\begin{array}{l}
s_{\rho} \equiv\langle r\rangle_{\rho}=\operatorname{Tr}[r \rho],\\\\ 
V_{\rho} \equiv\left\langle\left\{r-s_{\rho}, r^{T}-s_{\rho}^{T}\right\}\right\rangle_{\rho}=\operatorname{Tr}\left[\left\{r-s_{\rho}, r^{T}-s_{\rho}^{T}\right\} \rho\right].
\end{array}
\end{equation}

Consequently, we can express any Gaussian Bosonic Quantum Channel by two $2n \times 2n$ matrices $X$ and $Y$ which act on the state in the following way \cite{serafini2019quantum}:
\begin{equation} \label{channelapplication}
\begin{array}{l}
s \longmapsto X s, \\
V \longmapsto X V X^{\top}+Y .
\end{array}
\end{equation}

 For a channel to be a valid quantum operation (i.e. completely positive trace-preserving map) $X,Y$ must satisfy:

\begin{equation} \label{eq:channelcondition1}
Y+i J \geq i X J X^{\top}.
\end{equation}

 We examine following Bosonic Gaussian Channels (See Table \ref{tab:channels}): non-unitary \cite{serafini2019quantum} --- classical mixing channels which incoherently add some noise to the system, attenuator and amplification channels characterising the state's interaction with the thermal environment; unitary \cite{RevModPhys.84.621} ---  beam splitter channels representing the action of a transformation $\exp \left[\theta\left(\hat{a}^{\dagger} \hat{b}-\hat{a} \hat{b}^{\dagger}\right)\right]$ and squeezing channels describing the transformation: $\exp \left[r\left(\hat{a} \hat{b}-\hat{a}^{\dagger} \hat{b}^{\dagger}\right) / 2\right]$. 

\begin{table*}
\caption{\label{tab:channels} Overview of the considered Bosonic Gaussian channels. For each of the channels we specify: defining matrices $X$ and $Y$, the range of used parameters and the number of lattice sites affected. For the classical mixing channel, for the remainder of our considerations, $Y$ matrix is normalized to be of unit norm unless explicitly stated otherwise. $\mathbb{Z} = \text{diag} (1,-1)$.}
\begin{ruledtabular}
\begin{tabular}{ccccc}

 Name & $X$ & $Y$&Parameters
&$\#$sites affected\\ \hline
 Classical mixing channel & $\mathbb{I}$ & $\ge 0$ & $Y$ a random matrix& any $\ge 1$ \\
 Attenuator channel&$\cos \theta \mathbb{I}_{2}$
 &$(\sin \theta)^{2} n_{\text {th }} \mathbb{I}_{2}$&$\theta \in [0,2 \pi ),  n_{th} \geq 1$&$1$\\
 Amplification channel&$\cosh r \mathbb{I}_{2} $&$ (\sinh r)^{2} n_{\text {th }} \mathbb{I}_{2}$
 &$r \in[0, \infty),n_{\text {th }} \geq 1$&$1$\\
 Beam splitter channel&$\left(\begin{array}{cc}
\sqrt{\tau} \mathbb{I}_{2} & \sqrt{1-\tau} \mathbb{I}_{2} \\
-\sqrt{1-\tau} \mathbb{I}_{2} & \sqrt{\tau} \mathbb{I}_{2}
\end{array}\right)$&$0$&$\tau=\cos ^{2} \theta \in[0,1],\theta \in [0,2 \pi )$&$2$\\
 Squeezing channel& $\left(\begin{array}{ll}\cosh r \mathbb{I}_{2} & \sinh r \mathbb{Z} \\ \sinh r \mathbb{Z} & \cosh r \mathbb{I}_2\end{array}\right)$&$0$&$r \in \mathbb{R}$ &$2$\\
\end{tabular}
\end{ruledtabular}
\end{table*}

We also study the class of Petz Recovery maps which reverse an action of a general bosonic \footnote{Petz map studies with implications for free field theories have been conducted  in a Gaussian QI framework for a \emph{fermionic} case in \cite{Swingle2019}.} quantum channel $\mathcal{N}$ acting on $\sigma$. Their form is given by  \cite{lami2018approximate}:

\begin{equation}
\begin{array}{l}
X_{P} \equiv \sqrt{I+\left(V_{\sigma} J\right)^{-2}} V_{\sigma} X^{T} \left( \sqrt{I+\left(J V_{\mathcal{N}(\sigma)}\right)^{-2}} \right)^{-1} V_{\mathcal{N}(\sigma)}^{-1}, \\

Y_{P} \equiv V_{\sigma}-X_{P} V_{\mathcal{N}(\sigma)} X_{P}^{T}, \quad \mathrm{ where:}\\\\

V_{\mathcal{N}(\sigma)}=X V_{\sigma} X^{T}+Y.
\end{array}
\end{equation}

Such a unique linear map $\mathcal{P}_{\sigma, \mathcal{N}}$ exists for any $\sigma$ and $\mathcal{N}$ s.t. $\mathcal{N}(\sigma)$ is a faithful state i.e. it satisfies the following relation:
\begin{equation}
V_{\mathcal{N}(\sigma)}+ i J>0.
\end{equation}

 Presented channels are instantiations of completely positive trace preserving maps. Their action is equivalent, via a Gaussian Stinespring dilation \cite{serafini2019quantum}, to an application of a Gaussian unitary operator acting on an extended system followed by a process of tracing out the appended environment. For example, in the case of attenuator and amplification channels the environment is a Gibbs thermal state with an average number of excitations equal to $N = (n_{\text{th}}-1)/2$. Since $N$ can be calculated using Bose statistics, the $n_{\text{th}}$ parameter is directly related to the temperature and the frequency of the environmental state as follows: $(n_{\text{th}}-1)/2 = N = 1 /(e^{\beta \omega}-1)$ \cite{serafini2019quantum}.
 
 Last but not least, we can express unitary time evolution operators as Bosonic Gaussian quantum channels. From the technical perspective it is a crucial point since this enables us to consider the Hamiltonian dynamics intercepted with localized in time channel applications (see Table \ref{tab:channels}) while working solely in a Gaussian QI framework. In consequence, we can explore dynamics of not only pure but also mixed states of regularized QFTs.
 
 \subsubsection{Subsystems description}
 
 Presented framework, by construction, also encompasses a possibility to study the system not only on a global scale but also on a subsystem level which becomes instrumental when investigating the local disturbances in QFTs.

Let us assume that we want to divide a whole system of $n=a+b$ bosons into two subsystems $A$ and $B$ consisting of $a$ and $b$ bosons, respectively. Without loss of generality, we further impose that indices from $1$ to $2a$ correspond to the subsystem A. Then, the covariance matrix of the subsystem $A$ is just the top-left part of the covariance matrix of the whole system.  Mathematically, we have:

\begin{equation}
\begin{aligned}
V_{i j}^{A} &=\operatorname{Tr}_{A}\left(\boldsymbol{\rho}_{A}\left[c_{i}, c_{j}\right]\right) \quad i, j=1, \ldots, 2 a \\
&=\operatorname{Tr}_{A}\left(\operatorname{Tr}_{B}(\boldsymbol{\rho})\left[c_{i}, c_{j}\right]\right)=\operatorname{Tr}_{A}\left(\operatorname{Tr}_{B}\left(\boldsymbol{\rho}\left[c_{i}, c_{j}\right]\right)\right) \\
&=\operatorname{Tr}\left(\boldsymbol{\rho}\left[c_{i}, c_{j}\right]\right)=V_{i, j},
\end{aligned}
\end{equation}

\noindent where $c_i$ is the $i$th canonical operator. In this calculation, it was possible to take $\left[c_{i}, c_{j}\right]$ under the trace over $B$ because these canonical operators act only on the subsystem $A$.

\subsubsection{Local vs global channel action} \label{localglobalaction}

Having obtained the information about the subsystem, now, we would like to act with a bosonic operation exclusively on the extracted state. Consider the system consisting of two subsystems $A$ and $B$ described by their vectors of first moments and covariance matrices:

\begin{equation}
s_{i n}=\left(\begin{array}{l}
s_{A} \\
s_{B}
\end{array}\right) \quad, \quad V_{i n}=\left(\begin{array}{cc}
V_{A} & V_{A B} \\
V_{A B}^{\top} & V_{B}
\end{array}\right)
\end{equation}

If we apply a channel locally on the subsystem $A$, it will affect not only the part of the matrix corresponding to subsystem $A$. The result of applying a channel defined by $X$ and $Y$ on the subsystem $A$ is:

\begin{equation}
s_{\text {out }}=\left(\begin{array}{c}
X s_{A} \\
s_{B}
\end{array}\right) \quad, \quad V_{\text {out }}=\left(\begin{array}{cc}
X V_{A} X^{\top}+Y & X V_{A B} \\
V_{A B}^{\top} X^{\top} & V_{B}
\end{array}\right).
\end{equation}

Hence, such a channel also influences the off-diagonal cross-terms that define correlations between subsystems $A$ and $B$.

More details of how to obtain numerical values of entropy and fidelity for a Bosonic state written in the covariance matrix formalism are presented in Appendix \ref{appendix:qi}.
 
\subsection{Investigated states} \label{sec:investigatedstates}

In the following, we introduce states of interest and describe how to rewrite them into the language of Bosonic Gaussian QI. These are a vacuum state and a thermal state of free QFT$_{1+1}$ \footnote{Index $1+1$ refers to one spatial and one temporal dimension on which the field theory is defined.} (which we also refer to as 1D harmonic chain) and TFD of two entangled copies of free QFT$_{1+1}$. In holographic CFTs, these states correspond to  empty space-time, a black hole and to two wormhole-connected black holes \cite{Hubeny_2015,Maldacena_2003}, respectively.

The first step is to regularize the Hamiltonian of the field theory (Section \ref{sec:fieldregularization}) and then the second step is to induce out-of-equilibrium dynamics to the system (Section \ref{dynamic}). 

\subsubsection{Field theory regularization} \label{sec:fieldregularization}

\noindent In this article, we consider free field theories with the following Hamiltonian:

\begin{equation} \label{eq_Hamiltonian}
    H=\int_{-\mathcal{L} / 2}^{\mathcal{L} / 2} \mathrm{~d} x\left(\frac{1}{2} \pi(x)^{2}+\frac{1}{2} m^{2} \phi(x)^{2}+\frac{1}{2}\left(\partial_{x} \phi(x)\right)^{2}\right),
\end{equation}

\noindent where $\phi$ is the field variable, $\pi$ is the conjugate momentum variable and $m$ denotes the free mass of the field \footnote{To obtain the Conformal Field Theory limit one has to impose $m \to 0$.}.  It is a QFT living in one temporal and one spatial dimension with size $\mathcal{L}$ and with periodic boundary conditions (more details regarding the choice of boundary conditions can be found in \ref{section:boundary}). To make the problem tractable numerically we introduce a regularization (the regularization procedure closely follows \cite{Chapman_2019}) of the field by introducing a spatial lattice\footnote{In 1D spatial case we also use the word \textit{chain} instead of \textit{lattice}.}. Let us assume that the lattice consists of $N$ sites. Then the lattice spacing becomes $\delta = \frac{\mathcal{L}}{N}$. For such a discretized version of the chosen QFT the Hamiltonian becomes:

\begin{equation}
    H=\sum_{i=1}^{N}\left(\frac{\delta}{2} P_{i}^{2}+\frac{m^{2}}{2 \delta} Q_{i}^{2}+\frac{1}{2 \delta^{3}}\left(Q_{i}-Q_{i+1}\right)^{2}\right),
\end{equation}
where $Q_i=\phi(x_i)\delta$ and $P_i= \pi(x_i)$.\\

\noindent We further implement Real Discrete Fourier Transform which allows us to rewrite the Hamiltonian in the momentum basis as:

\begin{equation} \label{eq_Hdecoupled}
H=\sum_{n=1}^{N}\left(\frac{\hat{P}_{n}^{2}}{2M}+\frac{1}{2} M \omega_{n}^{2} \hat{Q}_{n}^{2}\right), 
\end{equation}

\noindent where:

\begin{equation} \label{eq_basis}
\begin{array}{l}
\hat{Q}_{n}=\frac{\tilde{Q}_{n}+\tilde{Q}_{N-n}}{\sqrt{2}}, \hat{P}_{n}=\frac{\tilde{P}_{n}+\tilde{P}_{N-n}}{\sqrt{2}},\\
 \hat{Q}_{N-n}=\frac{\tilde{Q}_{n}-\tilde{Q}_{N-n}}{\sqrt{2} i}, \hat{P}_{N-n}=\frac{\tilde{P}_{n}-\tilde{P}_{N-n}}{\sqrt{2} i},\\
 \tilde{Q}_{n}=\frac{1}{\sqrt{N}} \sum_{a=1}^{N} e^{i 2 \pi n a} Q_{a}, \quad \tilde{P}_{n}=\frac{1}{\sqrt{N}} \sum_{a=1}^{N} e^{i 2 \pi n a} P_{a}
\end{array}
\end{equation}

\noindent and $M = \frac{1}{\delta}$ becomes the effective mass of the system while:

\begin{equation}
\omega_{n}=\left(m^{2}+4 \delta^{-2} \sin ^{2} \frac{\pi n}{N}\right)^{1 / 2}.
\end{equation}

\noindent We see that in the Fourier-Transformed basis the Hamiltonian becomes a sum of decoupled Hamiltonians of harmonic oscillators each with its own natural frequency $\omega_n$. It is straightforward to calculate the time evolution of any state in such a form since we simply just let each of the modes to time evolve with its own frequency. 

\subsubsection{Dynamics -- out of equilibrium time evolution} \label{dynamic}

In this section, we explain in more detail how we treat the case of a time-dependent setup. Since evolution of QI measures in non-equilibrium dynamics settings is of much importance to the field \cite{Polkovnikov_2011, Eisert_2015,abanin2019colloquium, PhysRevLett.109.115304,PhysRevLett.98.070201,CAPUTA201753,Coser_2014} we also consider such dynamics in our work. 

If $t_0$ is the time of the quench application and the Hamiltonian transitions from $H_0$ to $H_1$, then we can characterize the evolution of the state as follows:
 
\begin{equation}
\ket{\psi(t)}=\left\{\begin{array}{ll}
e^{-iH_{1}(t-t_0)} \ket{\psi(t_0)} \mkern9mu \mathrm{for} \mkern9mu t \ge t_0\\
e^{-iH_{0}(t-t_0)} \ket{\psi(t_0)} \mkern9mu \mathrm{for} \mkern9mu t < t_0 
\end{array}\right..
\end{equation}
 
We implement a global quench for the 1D harmonic chain system by rescaling all the frequencies of the normal modes of the field: 
 
 \begin{equation}
\omega_{n}=\alpha \cdot\left(m^{2}+4 \delta^{-2} \sin ^{2} \frac{\pi n}{N}\right)^{1 / 2}
\end{equation}
 
\noindent by some universal constant factor $\alpha$. In this study, we primarily focus on the application of local quantum channels. As such, we anticipate the evolution of entropy in our system to exhibit universal behavior, irrespective of the specific quench protocol employed. This assertion aligns with the findings of \cite{Chen_2022,Coser_2014,Caputa_2017,Di_Giulio_2021}, who report similar universal behavior after a global quantum quench in 1D free lattice models and free scalar field theory. Nonetheless, this is an emerging field of study, and concrete assertions regarding the universality of entropy evolution require substantiation from CFT calculations, which we earmark as a potential avenue for subsequent research.

To impose global quench for the TFD state (written in the form of energy eigenmodes $\ket{E_n}$ decomposition): 

\begin{equation}
    \begin{array}{lr}
         \left|\operatorname{TFD}\left(t_{L}, t_{R}\right)\right\rangle=  \\
         \frac{1}{\sqrt{Z_{\beta}}} \sum_{n} e^{-\beta E_{n} / 2} e^{-i E_{n}\left(t_{L}+t_{R}\right)}\left|E_{n}\right\rangle_{L}\left|E_{n}\right\rangle_{R}, 
    \end{array}
    \end{equation}

\noindent we choose $t_L=t_R=t$ (the equilibrium situation would be represented by the choice $t_L=-t_R=t$ \cite{Hartman:2013qma, PhysRevD.90.126007}). $Z_{\beta}$ is the canonical partition function of the system.

\subsection{Channels standardization procedure} \label{sec:standardization}

\begin{figure*}
\includegraphics[trim={0 2.0cm 0 0},clip,scale=1.85]{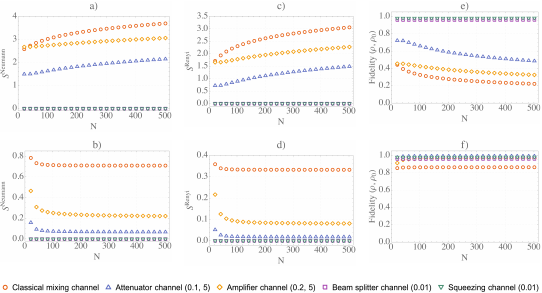}%
\caption{\label{fig:grid1} Figure presents how von Neumann entropy (a, b), R\'enyi entropy (c,d) and fidelity (e,f) change after the channel application --- their values are plotted as functions of the number of lattice sites $N$. The plots (a, c, e) show the results before the standardization of non-unitary Gaussian channels and (b, d, f) after the undertaken procedure. We observe that the standardization process fixes the convergence issues and allows us to define an action of non-unitary Gaussian channel in the continuum limit. The plots are obtained for $m=0.001/\mathcal{L}$ and R\'enyi entropy of order $5$. We calculate the entropies and fidelities for the full length harmonic 1D chain.}
\end{figure*}

The goal of the article is to combine the language of quantum channels with the language of field theories. That is why we would like to have the possibility to describe the channel action in the continuum limit. 

In the following, we explain the obstacles that are encountered while evaluating actions of channels in the limit of infinitely many lattice sites, i.e. $N \rightarrow \infty$ and how they can be circumvented with the standardization procedure we propose. We demonstrate the problem and its resolution on the example of a vacuum state of the 1D harmonic chain system. 

We measure von Neumman entropy and R\'enyi entropy for the whole system of the 1D harmonic chain after the channel application. We also calculate the fidelity between the excited state and the initial state of the full-length chain. We perform this analysis for all channels introduced in Section \ref{sec:channels}. We observe (see Figure \ref{fig:grid1}) that for non-unitary channels von Neumann entropy (Figure \ref{fig:grid1}(a)) as well as the R\'enyi entropy (Figure \ref{fig:grid1}(c)) diverge logarithmically with $N$ increasing as we apply the same channel on one particular site in the chain. Whereas, fidelity with respect to the original state for non-unitary channels approaches $0$ as  $N \rightarrow \infty$ (Figure \ref{fig:grid1}(e)).

On the other hand, for unitary Bosonic Gaussian channels, we notice very different trends. The evaluated entropies stay constant and equal to zero directly because of the unitarity property. Unitary channels do not mix the state and the state is initially pure. For the fidelity, we observe constant values with respect to $N$. This analysis allows us to conclude that non-unitary channels as opposed to unitary channels require modifications in order to obtain convergent continuum limit results. 

 We present a standardization of non-unitary Gaussian channels in the following form:

\begin{equation}
\begin{array}{c}
X_{N} =  \frac{1}{\log(N)}\cdot X + \left(1 - \frac{1}{\log(N)} \right)\mathbb{I}, \\\\
Y_{N} = \frac{Y}{N} .
\end{array}
\end{equation}

For this particular proposal $X_{N} \rightarrow \mathbb{I}$ and $Y_{N} \rightarrow 0$, as $N \rightarrow \infty$. This resolution is motivated by the fact that the norm of the covariance matrix is invariant with respect to $N$. Hence, $Y_{N}$ should scale in such a way that its norm should be proportional to the norm of the subsystem it acts upon. For this choice of $Y_{N}$, the condition:

\begin{equation} \label{eq:channelcondition}
Y+i J \geq i X J X^{\top}.
\end{equation}

\noindent that each channel must satisfy, restricts the limit value of $X_{N}$. We choose $X_{\infty} = \mathbb{I}$. The scaling factors for $X_{N}$ series, for growing $N$, are changing slower as for the case of $Y_{N}$ series. This choice ensures that $X$ part of the channel has a non-zero effect in the continuum limit.

Such a choice of standardization procedure for non-unitary Bosonic Gaussian channels (justified further in Section \ref{sec:standarizationdiscussion})  guarantees well-defined, non-trivial effect on a state in the field theory limit as follows from the analysis of plots Figure \ref{fig:grid1}(b), Figure \ref{fig:grid1}(d), Figure \ref{fig:grid1}(f) for which we observe rapid and stable convergence.

\subsection{Simulations technical summary} \label{sec:technicalsummary}

The goal of this section is to describe all the technical steps of simulations performed in order to obtain the results presented in Section \ref{sec:technicalsummary}. The necessary points are in direct correspondence with the previous subsections of Methods and Methodology Section \ref{methods}.

In Section \ref{sec:channels}, we discussed various non-unitary and unitary Bosonic Gaussian quantum channels. In the remainder of the paper, we focus solely on the classical mixing channel. We chose this channel because non-unitary channels have not been considered in the context of QFTs before. Moreover, one of the parameters of this channel is the number of lattice sites (modes) it can affect. Hence, we can study both the point-wise channel action as well as one with spatial extent for this particular channel. 

In Section \ref{sec:investigatedstates}, we presented states that we chose for further investigations because of their universality and importance for holography. In the Results Section \ref{results}, we always state with which state we will be working and why.

In Section \ref{sec:standardization}, we established a standardization procedure that allows us to work in the continuum limit. We always employ this procedure for each of the experiments. 

We have already discussed what channel and which states we take into consideration. Let us now explain how we investigate the chosen systems.

For the static case, before the channel application, we measure the QI measures using formulas from Section \ref{appendix:qi} applied to the covariance matrix (for more details see Section \ref{sec:channels}) of the state of interest. To calculate the needed covariance matrix we employ the Fourier-transformed basis (Equation \ref{eq_basis}) in which the Hamiltonian is diagonal. Then we decide whether we want to apply the channel globally or locally (this step is explained in Section \ref{localglobalaction}) and if locally then where in the system. After the channel application (defined by Equation \ref{channelapplication}) we once again calculate QI measures of interest and compare them against each other. 

For the dynamic case, the difference is that we also take into account the non-equilibrium time evolution of the system. Details of chosen dynamics are presented in the Subsection \ref{dynamic}. The updated Hamiltonian that governs the chosen global quench dynamics is also diagonal in the chosen Fourier-transformed basis (Equation \ref{eq_basis}), hence the time-evolution boils down to matrix multiplication. 

In the following simulations, the chain (lattice) is composed of $N$ sites (2$N$ for a TFD state). Periodic boundary conditions are imposed. Further adjustable parameters of the setup are: mass of the field $m$, inverse temperature $\beta$, circumference $\mathcal{L}$ of the region on which the field is defined and parameters defining the form of employed channels (see Table \ref{tab:channels}).

\section{Results} \label{results}

In Methods and Methodology Section \ref{methods}, we chose the Bosonic Gaussian Quantum Information framework as a common ground between quantum channels theory and complex many-body systems theory. As a direct consequence, we are able to investigate the properties of the excitation introduced to chain and lattice coupled systems. Moreover, because of the established well-defined behaviour in the case of the number of chain and lattice sites going to infinity, we can understand what channel action means for the continuous systems.  We quantify the excitation characteristics utilizing the notion of entanglement entropy which measures the \textit{spread} in the Hilbert space of a quantum system between the two subsystems --- one for which the excitation was introduced and the other that remains in its initial state. 

Recently, unitary Gaussian channels, which do not require the standardization procedure, were studied in the language of Quantum Field Theories \cite{gaussianinQFT1, gaussianinQFT2}. Hence, in the remaining part of this paper, motivated to broaden the scope of applicability of quantum channels, we focus on non-unitary Gaussian channels. For concreteness and because of the possibility of applying this channel to arbitrary number of modes, we investigate an operational meaning of a classical mixing channel. 

We investigate the action of this channel using three different states that we refer to as: 1D harmonic chain system vacuum and thermal and the TFD state all of which were introduced in the Methods and Methodology Section \ref{methods}.

We additionally choose when to use time-dependent and when time-independent setting.

In the following, we evaluate, using motivated quantum information measures, firstly how quantum channel affects the state itself studying the time-independent scenarios and secondly what dynamics it induces on top of the standard quantum mechanical time evolution of the system. 

\subsection{Static channel action} \label{sec:static}

\begin{figure*}
\includegraphics[trim={0 0.02cm 0 0.1cm},clip,scale=1.84]{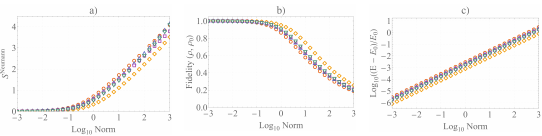}%
\caption{\label{fig:grid2}Figure presents how von Neumann entropy (a), fidelity (b) and energy (c) of the whole 1D harmonic chain system in a vacuum state changes after the application of classical mixing channel with varying norm. The dependencies are demonstrated for five random instantiations of a $Y$ matrix. $m=0.001/\mathcal{L}$ and $N=501$.}
\end{figure*}

To start, we study how an application of a channel changes the state of a system in a time-independent setting. 

\subsubsection{Strength of the channel analysis}
Firstly, we examine how an action of a single-mode classical mixing channel depends on the norm of its $Y_{1}$ matrix (Figure \ref{fig:grid2}) in the case of a vacuum state of 1D harmonic chain system. It means that we consider a channel that affects only one spatial site of the discretized system before we take the continuum limit. Hence, in the limit of infinitely many-sites this becomes a point-like excitation without any spatial extent. As an additional explanation, the norm of the classical mixing channel $Y_{1}$ directly corresponds to the amount of random uncorrelated information introduced into the system. This information can be viewed as created quasi-particles that distribute across the system. From this perspective, the norm of the channel defining the strength of the excitation corresponds directly to the number of quasi-particles created. Last but not least, the action of the classical mixing channel is fully determined by its $Y_{1}$ matrix, hence it is the parameter that has to be tweaked in order to understand the spectrum of possible excitations caused by this type of channel. 

We see (Figure \ref{fig:grid2}) that von Neumann entropy of the full length 1D harmonic chain exhibits a non-linear growth relation with respect to $\log_{10} (\mathrm{norm})$ with the increase rate advancing around $\mathrm{norm} = 1$. A similar breaking point for the derivative is obtained for the fidelity plot. The amount of disorder measured by the entanglement entropy and the amount of similarity to the initial state measured by fidelity are two independent indicators reflecting the level of deviation of the disturbed state from the initial state. The threshold above which these two show a significant change happens at $\mathrm{norm} = 1$. This is the point at which the entries of the $Y_N$ matrix are of the same order of magnitude as the entries to the covariance matrix of the system. Remember that both the matrix elements of the $Y_N$ matrix as well as the matrix elements of the covariance matrix are reduced in magnitude with growing number of sites $N$. Hence, the channel matrix has to be comparable to the quadrature elements of a single site of the lattice it acts upon to produce an effect visible globally in the system.

Lastly, for the energy-norm graph, we obtain a linear trend on the log-log scale. Since the slope of these lines is exactly $1$, we establish a captivating relation that the energy added by the classical mixing channel scales linearly with the norm. This is related to the fact that each single site in the lattice is treated as a simple harmonic oscillator that is coupled to the rest of the system. Its energy is directly proportional to the quadrature position and momentum elements expressed by the covariance matrix, hence boosting up these quadrature terms, causes the energy to grow with the same scaling.

\begin{figure}
\includegraphics[trim={0 0.1cm 0 0.2cm},clip,scale=0.50]{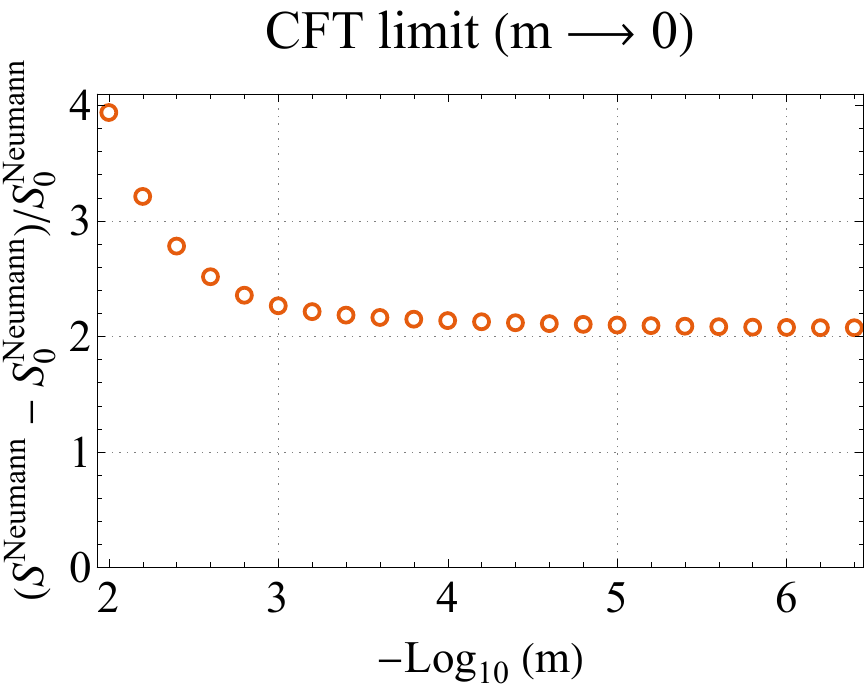}% Here is how to import EPS art
\caption{\label{fig:CFT} Figure shows the relative change in the entropy of a thermal state of a 1D harmonic chain system after the application of a classical mixing channel (introduced on one site of the lattice) as a function of scalar mass of the field. In the CFT limit we obtain a well-defined non-zero value of the measured change. $N=501$, $\mathcal{L} = 1$, $\beta = 75$ and $||Y_1||=50$.}
\end{figure}

\subsubsection{Massless field limit} \label{section:massless}
Secondly, we analyze the action of a classical mixing channel on a thermal state of a 1D harmonic chain system in the limit of vanishing mass of the field, $m \rightarrow 0$. We cannot brute force set the mass to vanish since the terms in the covariance matrix capturing the whole information about the system are inversely proportional in magnitude to that parameter. That causes the norm of the covariance matrix to diverge as $m$ approaches $0$.

As a result, for achieving clear-cut outcomes, it's essential that we evaluate our selected quantum information measures based on their relative changes rather than focusing on their absolute values.

Furthermore, we have to introduce additional regularization of the $Y$ matrix describing the strength of the excitation caused by the classical mixing channel. We impose that $Y_{N,m}= \frac{Y_1}{m^2 \cdot N}$ as a one possible regularization choice. For such a regularization, we obtain convergence of the relative change in the entropy of a thermal state of a 1D harmonic chain system (see Figure \ref{fig:CFT}).

We have already established the way of defining the action of classical mixing channel in the continuum limit. Combined with this result, we demonstrate a procedure that allows us to obtain parameter invariant results that reflect the exact characteristics of CFTs. This establishes a framework that allows for direct verification of hypotheses and arguments using theoretical approaches that have to resort to approximations. Not only it allows to challenge results obtained theoretically, but it provides a systematic method of quantifying which approximations have the most impact on the final result. 

\subsection{Dynamics of quantum information} \label{sec:QIdynamics}

In the previous subsection, we focused on the static action of the channel. We analysed the properties of the excitation formed by carefully inspecting the state before and after channel application. What remains to be understood is how the time dynamics of the state are influenced by such an excitation. How does the excitation propagate, what long-term effect we get when we try to undo a disturbance --- these are research questions we address in this section.

\subsubsection{Quasi-particle picture}

\begin{figure}
\includegraphics[trim={0 0.32cm 0 0.08cm},clip,scale=0.50]{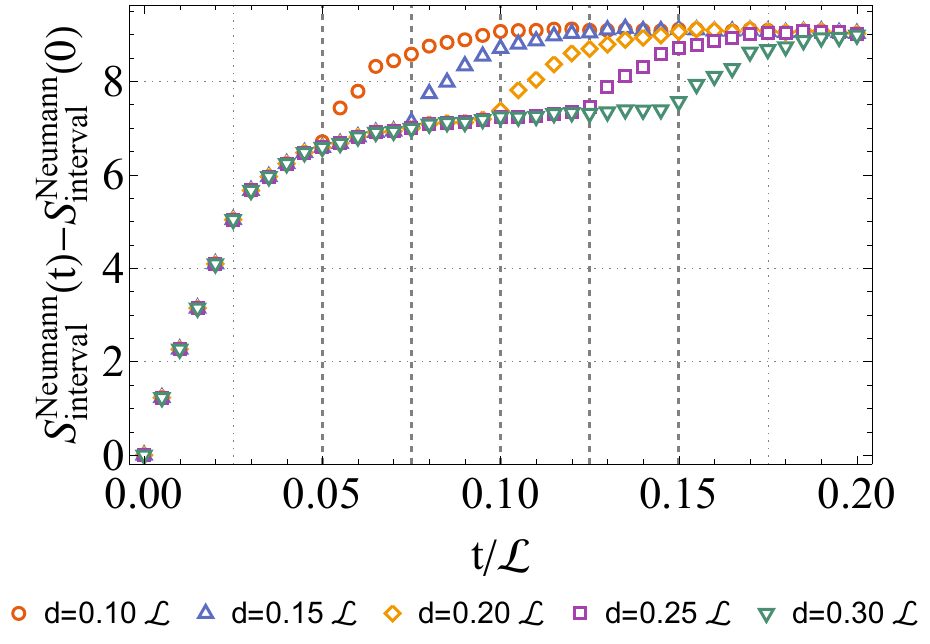}% Here is how to import EPS art
\caption{\label{fig:mobility} Figure presents the evolution in time of von Neumann entropy of an interval of length $l=0.1\mathcal{L}$ of a vacuum state of a 1D harmonic chain system. This evolution is disturbed by introducing classical mixing noise at $t=0$ in a distance $d$ from the chosen interval. A classical mixing channel acts on a single lattice site. Dashed vertical lines indicate when the excitations of the system caused by the perturbation reach the interval of interest. $m=0.001/\mathcal{L}$, $N=201$, quench factor $\alpha = 2.0$ and $||Y_1||=50$.}
\end{figure}

To begin with, we evaluate the dynamics of quantum information in the context of a 1D harmonic chain system in a vacuum state with the evolution disturbed by a classical mixing channel. We examine how von Neumann entropy of an interval of length $l=0.1\mathcal{L}$ varies in time (see Figure \ref{fig:mobility}). This evolution is perturbed, at $t=0$, by classical mixing noise in a distance $d$ from the interval of interest. At first, the entanglement entropy increases linearly and then saturates at a level proportional to the size of the subsystem $l$ \cite{Coser_2014}. In addition, we notice that for all the values of $d$ the calculated relative change in von Neumann entropy saturates at the same value, thereby respecting the translation symmetry of the system. Later on, the entanglement entropy offsets when the lattice excitation induced by the channel arrives at the considered interval. We observe that for equally spaced values of $d$ the entropy elevates at equally spaced moments in time. This phenomenon is consistent with the quasi-particle picture, which was introduced to understand the entanglement spreading in TFD states \cite{Calabrese:2005in, Calabrese:2016xau,Chapman_2019}. In such a framework, we treat quasi-particles as excitations of the Hamiltonian normal modes spreading across the system at constant velocities $v_n$ that, in the continuum limit $N \rightarrow \infty$, are given by:

\begin{equation}
v_n =  \frac{\mathcal{L}}{2 \pi} \frac{\partial }{\partial n} \left(\lim_{N\to\infty} \omega_n\right).
\end{equation}

To obtain the velocity of the forefront of the disturbance we maximize $v_n$ with respect to $n$, and get $v_{\mathrm{max}} = \alpha$.

Therefore, in our case, $v_{\mathrm{max}} = \alpha = 2$, which is in agreement with results in Figure \ref{fig:mobility}. Hence, we conclude that the aforementioned analysis provides a strong indication that excitations caused by a classical mixing channel (as well as possibly by other non-unitary and unitary Gaussian channels) can be described in the quasi-particle framework.

\begin{figure}
\includegraphics[trim={0 0.47cm 0 0.3cm},clip,scale=0.59]{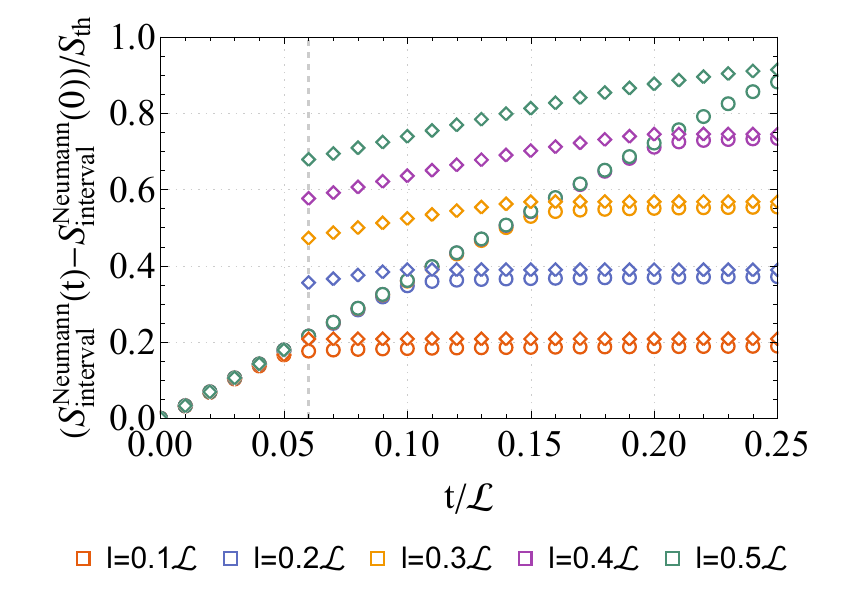}% Here is how to import EPS art
\caption{\label{fig:TFD} Figure presents the entanglement entropy growth for the free evolution (circles) and the evolution abruptly perturbed (diamonds) at $t=0.06$ by a classical mixing channel affecting all the sites on the left of the TFD state promptly succeeded by Petz Recovery map corresponding to the introduced noise. We calculate the difference in von Neumann entropy with respect to the initial state for the subsystem consisting of two identical intervals one on each side of TFD, both of length $l$ (different $l$ values are represented by different colours in the plot). We normalize this quantity by the entropy $S_{th}$ of a corresponding thermal state of size $N$. $m=0.001/\mathcal{L}$, $\beta = 0.01 \mathcal{L}$, $2N=1002$ and $||Y_1||=50$.}
\end{figure}

\subsubsection{Long-term effect of Petz recovery map}

To invoke more complicated time evolution dynamics, we refer to the TFD state. It consists of two QFTs entangled with each other, hence the state is defined not only by the state of each of the QFTs but also by the entanglement interconnections between them which cause the correlations to appear in such a setup. For the sake of simplicity let us call the marginal QFTs states left and right. 

Let us consider the following scenario. We measure, during the time evolution, the entanglement entropy of a subsystem consisting of an interval of length $l$ on the left side of TFD and of the corresponding interval of length $l$ on the right side. The characteristics of such a subsystem are defined by the entanglement interaction between the two sides of the TFD. Specifically, we study the entanglement entropy growth in the setting of free evolution \cite{Chapman_2019} and the evolution disturbed at $t =0.06$ by the classical mixing channel applied on the whole left side of TFD and immediately followed by Petz Recovery map perturbation also applied to whole left side (see Figure \ref{fig:TFD}). Such a combination of channels leaves both sides of the TFD state unchanged, however, it modifies the cross-terms of covariance matrix defining the entanglement between the interconnected regularized QFTs. Namely, the restrictions of this state either to the left or to the right side, obtained via a trace operation, are not altered; however, the correlations between canonical operators from different sides are modified.

As one of the main findings of this work, we discover that after such a procedure, the entanglement entropy still saturates at the same level; however, immediately after the operation, it takes a value in proximity to the plateau, which indicates that the system has thermalized. Hence, in the long run no sign of introduced channel action prevails. The long-term effect of the Petz recovery map is to perfectly recover the system, although the entanglement terms were altered. This alteration manifests itself only just after the channel application via faster thermalization, but the final effect is the same.

\section{Discussion} \label{discussion}

\subsection{Related work and results importance} \label{relatedworkandresults}

\subsubsection{Context and background for this work} \label{context}

Quantum Information Theory has become an extensive toolkit for modern mathematical and theoretical physics. One of its inherent features is its operational character that allows gaining new perspectives on scenarios previously only discussed using non-operational theories. A recent summary of progress of applying QIT methods to QFT domain (see \cite{QFTQIT}) discusses how entanglement entropy, information spreading and the role of information are indispensable in ongoing investigations. An article \cite{Chen_2022} takes it one step further and analyses how and why R\'enyi entropies, modular minimal entropy and entanglement wedge are essential concepts for AdS/CFT and holographic duality studies. 

Behaviour of quantum information measures has already been studied in non-equilibrium quantum dynamics settings \cite{Polkovnikov_2011, Eisert_2015,abanin2019colloquium, PhysRevLett.109.115304,PhysRevLett.98.070201,CAPUTA201753,Coser_2014} as well as in the context of holography \cite{Chapman_2019}. These advances directly fall into the category of utilizing QIT ideas to gain new understanding of previously established concepts. 

Moreover, not only have quantum information measures been used, but also operations and transformations native to QIT have been introduced with great success to holography studies, particularly in the entanglement wedge reconstruction and in resolutions of the black hole information paradox \cite{Cotler:2017erl,penington2020entanglement, Chen:2021lnq}.

There was, however, still an unexplored territory regarding understanding what role quantum channels can play in QFT, CFT, and holography. Both the methods and means of evaluating information dynamics have been previously established but eventually only this work created a suitable framework to determine what it would mean to apply a channel to a field theory --- whether it might be treated as an excitation, if yes of what kind and characteristics and what information dynamics it causes in the system. Hence, the key player of this work is a quantum channel, especially a Bosonic Gaussian channel which action has not been previously discussed in QFT, CFT, neither in holography context. 

\subsubsection{New findings and created insights}

Summarising our efforts: as our model systems, we investigated a vacuum and a thermal state of a 1D QFT and a Thermofield Double state that are universal and play a significant role in many-body physics and holography studies. We analysed and quantified actions of Bosonic Gaussian channels and their Petz Recovery maps for the proposed states both in a time-dependent and a time-independent setting. We considered unitary and non-unitary Bosonic Gaussian channels with emphasis put on a classical mixing channel.

All of the above leads to the creation of a new QI framework that allows studying quantum channels as excitations in QFT and CFT. Although, as outlined in the previous Section \ref{context}, efforts to combine QI and QFT studies are ubiquitous, quantum channels operations were never investigated in the context present in this work. The novelty of the framework was to find the common ground in the form of Bosonic Gaussian QI that combines the operational character of QI and continuous character of field theories. Proposing such a framework is already a significant addition to the field since it enables research at an intersection not yet explored.

The framework allows for the following: firstly, one can apply any Bosonic Gaussian channel in the form of a point-like excitation, as well as an excitation with spatial extent, to any field theory of their choice, including conformal field theories; secondly, it is possible to quantify and evaluate the action of the introduced channel excitation using quantum information measures such as Von Neumann entanglement entropy, R\'enyi entanglement entropy, and fidelity; lastly, one might determine the dynamics of induced quantum information in the setup of their choice where such dynamic has already been verified to be in agreement with previously established models and understanding.

\subsection{Impact of boundary conditions} \label{section:boundary}

Throughout this paper, our primary focus has been on systems with periodic boundary conditions, also referred to as von Neumann boundary conditions. This choice, as opposed to the alternative of employing Dirichlet boundary conditions, was driven by the profound implications that these boundary conditions have on the topology, symmetry, and quantum field theory (QFT) properties of the system.

Periodic boundary conditions, which render the system topologically equivalent to a circle (in 1D) or a torus (in 2D), simplify the analysis due to inherent symmetries. This leads to more manageable mathematical formulations and unveils intriguing topological phenomena.

Unlike Dirichlet boundary conditions, periodic ones do not introduce artificial boundaries, which can influence the system's behavior. This is particularly relevant in quantum information studies, where boundary effects can significantly alter the system's entanglement properties \cite{Berthiere_2019}.

Moreover, especially for lattice QFTs, periodic boundary conditions help minimize finite size effects, thereby providing a more accurate representation of infinite systems \cite{Solodukhin_1999}.

On the other hand, transitioning to Dirichlet boundary conditions in a 1D harmonic chain would imply that the field vanishes at the boundaries, effectively confining the system \cite{choi2023comments}. This confinement leads to a more localized response of a locally applied quantum channel, contrasting with the unrestricted propagation of effects in the case of periodic boundary conditions \cite{Park_2017}.

The imposition of Dirichlet boundary conditions can also alter the spectral properties of the system \cite{morley2020quantum}. For instance, the presence of zero modes, which can lead to infrared divergences in the case of periodic boundary conditions, is avoided under Dirichlet boundary conditions. Consequently, the further standardization procedure relative to the free mass of the field, as detailed in Section \ref{section:massless}, becomes unnecessary with this choice of boundary conditions.

While the choice of boundary conditions can influence the behavior of a system, our results remain largely consistent across both Dirichlet and periodic boundary conditions. The core findings of our study, as presented in this paper, are not significantly affected by these changes. This is in line with the findings of \cite{Di_Giulio_2019, Guo_2015}, who also reported minimal influence of the boundary conditions on the entanglement properties of 1D free lattice models and 2D CFTs, respectively. 

\subsection{Channel standardization procedure} \label{sec:standarizationdiscussion}

The standardization procedure for non-unitary Gaussian channels, as described in Section \ref{sec:standardization}, is further justified when we look at the wider context of quantum information theory related to quantum channels.

In \cite{Gyongyosi_2014}, the idea of partially degradable (PD) quantum channels is introduced. Here, the output state of the channel can simulate the degraded environment state. The quantum capacity of a PD channel is shown to be additive. This suggests that our standardization procedure could be seen as a kind of partial degradability.

Also, survey \cite{Gyongyosi_2018} on quantum channel capacities gives a full overview of the properties of quantum communication channels and the different capacity measures. Our standardization procedure can be seen as a way to optimize channel capacity. This is done by adjusting the scaling of the operators $X_N$ and $Y_N$ based on the number of lattice sites $N$. This ensures that the action of the channel in the continuum limit is well-defined and non-trivial.

\subsection{Characterisation of classical mixing channel in the density operator formalism} \label{appendix:gaussian}

In this section, we indicate a possibility how one could understand actions of quantum channels in QFTs from the analytical perspective. This is the first step towards opening a broad research avenue that we would like to only get a grasp of here. That is because the focus of this paper was to introduce the presented framework and demonstrate its inner-working by tangible numerical simulations. 

Accordingly, let us consider here a simplistic setup and at the end of this analysis we indicate how calculations and reasoning have potential to be extended further to a much broader class of systems. 

Let us start with a single quantum harmonic oscillator being in its ground state, i.e. $\sigma = \ket{0}\bra{0}$. For this state, we can easily calculate the covariance matrix before and after the application of the classical mixing channel ($s_{\sigma} = 0$ for $\ket{0}$):

\begin{equation}
\mathcal{N}: V_{\sigma_0} = \left(\begin{array}{ll} 1 & 0 \\ 0 & 1 \end{array}\right)  \longmapsto V_{\sigma} =  \left(\begin{array}{ll} 1 & 0 \\ 0 & 1 \end{array} \right) + Y.
\end{equation}

Operations on covariance matrices, although mathematically compact and convenient, leave us within the quantum information finite systems framework. However, it is possible to go back from this formalism to the operator representation of the state via the following integral \cite{serafini2019quantum}:

\begin{equation}
\sigma=\frac{1}{(2 \pi)^{n}} \int_{\mathbb{R}^{2 n}} \mathrm{~d} \mathbf{r} \quad \mathrm{e}^{-\frac{1}{4} \mathbf{r}^{\top} J^{\top} V_{\sigma} J \mathbf{r}}  \hat{D}_{\mathbf{r}}
\end{equation}

\noindent where: $\hat{D}_{\mathbf{r}}  =e^{i \mathbf{r}^{\top} J \hat{r}}$.

\begin{figure}
\includegraphics[scale=0.9]{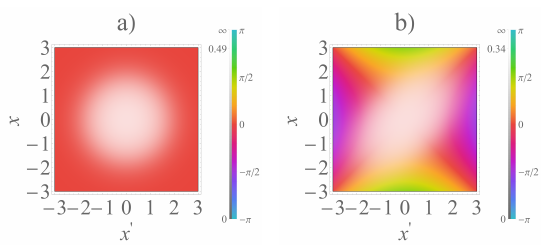}% Here is how to import EPS art
\caption{\label{fig:grid3}Position representation of the density operator for a vacuum state of a single simple harmonic oscillator before (a) and after (b) the application of a classical mixing channel. The brightness indicates the absolute value and color the phase of a complex number at a given point.}
\end{figure}

Following such a procedure, we can obtain the position representation of the density operator for a vacuum state $\ket{0}$ perturbed with the classical mixing channel (see Figure \ref{fig:grid3}). We find that the coherences i.e. the off-diagonal terms $\rho(x',x)$ such that $x' \neq x$ are squeezed by the channel. The norm of the state gets concentrated near the $x=x'$ line for which the density operator represents the classical probability of finding the particle at the position $x=x'$. In addition to the compression effect, we note that after the channel application, a non-zero phase appears across the whole state. The action of other non-unitary channels also results in pressing the state towards the $x=x'$ line but does not introduce any phase dependency across it. Analogous reasoning also applies and was carried out for the TFD of two quantum harmonic oscillators. Observed effects possess the same qualitative characteristics as for a vacuum state but it is not possible to visualize them in two dimensions.

It is crucial to note that the conducted analytical analysis can be extended to an arbitrary number of quantum modes in the system because the emerging integrals are of Gaussian type and hence can be evaluated analytically for arbitrarily high $N$. We believe that this analysis might be a starting point for the analytical calculations in the continuum limit which eventually we would like to reach following the initiated research direction. 

\section{Conclusions} \label{conclusions}

We have established a novel Quantum Information framework that allows for systematic studies of excitations in QFTs as quantum channels.
 
Moreover, we proposed a standardization of non-unitary channels, hence providing a way to obtain well-defined results in the field theory continuum limit. 

We characterized the spectrum of possible excitations induced by a channel by studying how the quantum information measures change with the channel defining parameters. 

In addition to that, we presented an approach for quantifying the action of a channel in the CFT limit of $m \rightarrow 0$. Hence, we paved the way towards predicting the intrinsic CFT quantities working in the proposed QI framework. 
  
We studied, furthermore, the dynamics of excitations caused by quantum channels proving their compatibility with the quasi-particles picture. 

Finally, we found that noise application followed by immediate Petz Recovery map does not pose any long-term effects, it only causes the system to thermalize in a more abrupt manner. Hence, the action of a channel can be reversed also in the dynamical setting. 
  
Our future goal is to understand physical meaning of non-unitary Bosonic Gaussian channels in the language of CFT and also in the context of holography. We aim to realize the introduced notions from a rigorous analytical point of view, by taking a similar line of reasoning as in \cite{Nozaki2014QuantumEO}. We already indicate the key lines of reasoning to be taken into account into that process in the Discussion Section \ref{appendix:gaussian}.

We hope that the proposed operational framework will become a tool in the systematic studies of excitations in QFTs and CFTs. \\

\begin{acknowledgements}
We acknowledge advice from Paweł Caputa, Mischa Woods and Renato Renner. We thank Michał Heller for comments regarding implementation and for providing us with a code that we based upon, Mark Wilde for helpful explanations regarding Gaussian QIT and Lucas Hackl for clarifications shared via e-mail correspondence.  M.B. is supported by NCN Sonata Bis 9 grant.\\
  
\end{acknowledgements}

\newpage

\appendix

\section{Bosonic Gaussian Quantum Information} \label{appendix:qi}

\noindent Throughout the paper we refer to calculations of various quantum information measures in the covariance matrix formalism. We provide, here, the technical details of these computations. \\

\noindent \textbf{Von Neumann entropy.} Consider the following function:

\begin{equation}
s(\lambda)=\left(\frac{\lambda+1}{2}\right) \log \left(\frac{\lambda+1}{2}\right)-\left(\frac{\lambda-1}{2}\right) \log \left(\frac{\lambda-1}{2}\right),
\end{equation}

\noindent which is defined on $[1, \infty)$ and takes values in the range $[0, \infty)$. Then von Neumann entropy of a $n$ mode Gaussian state $\rho(t)$ with a vector of first moments equal to $0$ and a covariance matrix $V(t)$ is \cite{Chapman_2019}:

\begin{equation}
S\left(\rho(t)\right)=\frac{1}{2} \sum_{i} s\left(\left|\lambda_{i}\right|\right),
\end{equation}

\noindent where:

\begin{equation}
\lambda_{i} \text {s are eigenvalues of } V^{1 / 2}(t)\left(i J_{n}\right) V^{1 / 2}(t).
\end{equation}

\noindent \textbf{R\'enyi entropy.} We define $\lambda_{i}$s as in the case of Von Neumann entropy. Then R\'enyi entropy of a $n$ mode Gaussian state $\rho(t)$ with a vector of first moments equal to $0$ and a covariance matrix $V(t)$ is
\cite{Chapman_2019}:

\begin{equation}
S_{q}\left(\rho(t)\right)=\frac{1}{2} \sum_{i} s_{q}\left(\left|\lambda_{i}\right|\right),
\end{equation}

\noindent where $s_q$ function is parameterized by a parameter $q>0$ and takes the following form:

\begin{equation}
s_{q}(\lambda)=\frac{1}{q-1} \log \left[\frac{(\lambda+1)^{q}-(\lambda-1)^{q}}{2^{q}}\right].
\end{equation}

\noindent \textbf{Fidelity.} Fidelity is a measure of distance between two quantum states and in the case of two Gaussian states $\rho_1$ and $\rho_2$ (each of $n$ bosons) with vectors of first moments equal to $0$ and covariance matrices $V_1$ and $V_2$ it can be expressed as \cite{banchi2015quantum}:  

\begin{equation}
\mathcal{F}\left(\rho_{1}, \rho_{2}\right)=\mathcal{F}_{0}\left(V_{1}, V_{2}\right),
\end{equation}

\noindent where:

\begin{equation}
\begin{aligned}
\mathcal{F}_{0}\left(V_{1}, V_{2}\right) &=\frac{F_{\mathrm{tot}}}{\sqrt[4]{\operatorname{det}\left[ \frac{V_{1}+V_{2}}{2}\right]}}, \\
F_{\mathrm{tot}}^{4} &=\operatorname{det}\left[2\left(\sqrt{\mathbb{I}+\frac{\left(V_{\mathrm{aux}} J_{n}\right)^{-2}}{4}}+\mathbb{I}\right) V_{\mathrm{aux}}\right] \\
&=\operatorname{det}\left[\left(\sqrt{\mathbb{I}-W_{\mathrm{aux}}^{-2}}+\mathbb{I}\right) W_{\mathrm{aux}} i J_{n}\right]
\end{aligned}
\end{equation}

\noindent and:

\begin{equation}
\begin{array}{l}
V_{\mathrm{aux}}=J_{n}^{T}\left(\frac{V_{1}+V_{2}}{2}\right)^{-1}\left(\frac{J_{n}}{4}+\frac{V_{2}}{2} J_{n} \frac{V_{1}}{2}\right) \\
W_{\text {aux }}:=-2 V_{\text {aux }} i J_{n} .
\end{array}
\end{equation}

\bibliography{References}% Produces the bibliography via BibTeX.

\end{document}